\documentclass[]{emulateapj}
\usepackage{apjfonts}
\usepackage{natbib}
\usepackage{epsfig}
\usepackage{amsmath}
\shorttitle{CHARACTERIZATION OF CENTAURUS A WITH QUAD}
\shortauthors{ZEMCOV ET AL.}

\begin{document}

\slugcomment{Received 2009 August 17; accepted 2009 December 30; published 2010 February 1}

\title{Characterization of the Millimeter-Wave Polarization of Centaurus A
with QUaD}

\author{
  M.\,Zemcov\altaffilmark{1,2,3},
  P.\,Ade\altaffilmark{3},
  J.\,Bock\altaffilmark{1,2},
  M.\,Bowden\altaffilmark{3,4},
  M.\,L.\,Brown\altaffilmark{5},
  G.\,Cahill\altaffilmark{6},
  P.\,G.\,Castro\altaffilmark{7},
  S.\,Church\altaffilmark{4},
  T.\,Culverhouse\altaffilmark{8},
  R.\,B.\,Friedman\altaffilmark{8},
  K.\,Ganga\altaffilmark{9},
  W.\,K.\,Gear\altaffilmark{3},
  S.\,Gupta\altaffilmark{3},
  J.\,Hinderks\altaffilmark{4,11},
  J.\,Kovac\altaffilmark{2},
  A.\,E.\,Lange\altaffilmark{2},
  E.\,Leitch\altaffilmark{1,2},
  S.\,J.\,Melhuish\altaffilmark{3,12},
  Y.\,Memari\altaffilmark{10},
  J.\,A.\,Murphy\altaffilmark{6},
  A.\,Orlando\altaffilmark{2,3}
  C.\,O'\,Sullivan\altaffilmark{6},
  L.\,Piccirillo\altaffilmark{3,12},
  C.\,Pryke\altaffilmark{8},
  N.\,Rajguru\altaffilmark{3,13},
  B.\,Rusholme\altaffilmark{4,14},
  R.\,Schwarz\altaffilmark{8},
  A.\,N.\,Taylor\altaffilmark{10},
  K.\,L.\,Thompson\altaffilmark{4},
  A.\,H.\,Turner\altaffilmark{3},
  and
  E.\,Y.\,S.\,Wu\altaffilmark{4}
  (QUaD collaboration)
}

\altaffiltext{1}{Jet Propulsion Laboratory, 4800 Oak Grove Dr.,
  Pasadena, CA 91109, USA.}
\altaffiltext{2}{California Institute of Technology, Pasadena, CA
  91125, USA.}
\altaffiltext{3}{School of Physics and Astronomy, Cardiff University,
  Queen's Buildings, The Parade, Cardiff CF24 3AA, UK.}
\altaffiltext{4}{Kavli Institute for Particle Astrophysics and
Cosmology and Department of Physics, Stanford University,
382 Via Pueblo Mall, Stanford, CA 94305, USA.}
\altaffiltext{5}{Kavli Institute for Cosmology and Cavendish Astrophysics,
  University of Cambridge, Madingley Road, Cambridge CB3 OHA, UK.}
\altaffiltext{6}{Department of Experimental Physics,
  National University of Ireland Maynooth, Maynooth, Co. Kildare,
  Ireland.}
\altaffiltext{7}{CENTRA, Departamento de F\'{\i}sica,
  Edif\'{\i}cio Ci\^{e}ncia, Piso 4,
  Instituto Superior T\'ecnico - IST, Universidade T\'ecnica de Lisboa,
  Av. Rovisco Pais 1, 1049-001 Lisboa, Portugal.}
\altaffiltext{8}{Kavli Institute for Cosmological Physics,
  Department of Astronomy \& Astrophysics, Enrico Fermi Institute,
  University of Chicago, 5640 South Ellis Avenue, Chicago, IL 60637, USA.}
\altaffiltext{9}{APC/Universit\'e Paris 7 - Denis
  Diderot/CNRS, B\^atiment Condorcet, 10, rue Alice Domon et L\'eonie
  Duquet, 75205 Paris Cedex 13, France}
\altaffiltext{10}{Institute for Astronomy, University of Edinburgh,
  Royal Observatory, Blackford Hill, Edinburgh EH9 3HJ, UK.}
\altaffiltext{11}{Current address: NASA Goddard Space Flight
  Center, 8800 Greenbelt Road, Greenbelt, MD 20771, USA.}
\altaffiltext{12}{Current address: School of Physics and Astronomy,
  University of Manchester, Manchester M13 9PL, UK.}  
\altaffiltext{13}{Current address: Department of Physics and
  Astronomy, University College London, Gower Street, London WC1E 6BT, UK.}
\altaffiltext{14}{Current address:
  Infrared Processing and Analysis Center,
  California Institute of Technology, Pasadena, CA 91125, USA.}

\begin{abstract}
Centaurus (Cen) A represents one of the best candidates for an
isolated, compact, highly polarized source that is bright at typical
cosmic microwave background (CMB) experiment frequencies.  We present
measurements of the $4^{\circ} \!  \times 2^{\circ}$ region centered
on Cen A with QUaD, a CMB polarimeter whose absolute polarization
angle is known to $0^{\circ}\!\!\!.5$.  Simulations are performed to assess
the effect of misestimation of the instrumental parameters on the
final measurement, and systematic errors due to the field's background
structure and temporal variability from Cen A's nuclear region are
determined.  The total $(Q, U)$ of the inner lobe region is $(1.00 \pm
0.07 \, (\mathrm{stat.}) \pm 0.04 \, (\mathrm{sys.}), -1.72 \pm 0.06 \pm
0.05)\,$Jy at 100 GHz and $(0.80 \pm 0.06 \pm 0.06, -1.40 \pm 0.07 \pm
0.08)\,$Jy at 150 GHz, leading to polarization angles and total errors
of $-30^{\circ}\!\!\!.0 \pm 1^{\circ}\!\!\!.1$ and $-29^{\circ}\!\!\!.1 \pm
1^{\circ}\!\!\!.7$.  These measurements will allow the use of Cen A as a
polarized calibration source for future millimeter experiments.
\end{abstract}


\vspace*{10pt}

\section{Introduction}
\label{S:intro}

\setcounter{footnote}{0}

Cosmic microwave background (CMB) polarimeters require accurate
calibration of both their temperature and polarization properties.
Unfortunately, astronomical sources suitable for polarization
calibration in the wavelength range at which CMB experiments typically
operate are rare.  With a number of exquisitely sensitive instruments
in the field or near to deployment, the need for astrophysical
polarization calibration sources useful for instruments with beam
sizes of several to tens of arcmin is acute.

An ideal polarized calibration source would be compact, bright in both
temperature and polarization, static over time, and isolated from
galactic diffuse emission.  Unfortunately, the typical frequencies at
which CMB experiments are designed to measure are minima in the
emission spectra for other astrophysical sources of radiation
(synchrotron, free--free, and thermal dust emission), so such sources
have not been identified in large numbers or studied in great detail
at the frequencies of interest.  A noteworthy candidate calibration
source which has been observed with CMB polarimeters in the past is the
supernova remnant Taurus A (\citealt{Leitch2002},
\citealt{Barkats2005}), though as an object in the northern
hemisphere it is not available to telescopes at all sites.

One of the best candidates for a CMB polarization calibration
source in the southern hemisphere is the radio bright galaxy Centaurus
A (hereafter Cen A; \citealt{Israel1998} provides an excellent review
of its multi-wavelength properties).  The optical counterpart of Cen A
is the massive elliptical galaxy NGC 5128 at a distance of $3.4
\,$Mpc.  At radio frequencies Cen A presents rich structure over many
decades in angular size driven by its nuclear source.  Collimated
radio jets are emitted from the compact nucleus and become sub-sonic a
few parsec ($\sim 0.1"$) from the central source.  At 5 kpc ($\sim
4'$) from the nucleus, the jets expand into plumes which spread up to
250 kpc ($\sim 3^{\circ}$) into the inter cluster medium (ICM).  The
interface between the bright sub-sonic jets and the smooth, low
surface brightness plumes is abrupt, allowing high signal to noise
measurements of the jets themselves.  The $\sim 10'$ jets, known as
Cen A's `inner lobes' in the literature, have a spectral index of
$-0.7$ from $> 500 \,$MHz to $\sim 100 \,$GHz and are both polarized
and bright at $\nu \gtrsim 100 \,$GHz.  Moreover, Cen A's inner lobes
are unique as their properties do not vary on human time scales, which
makes them perfect candidates for polarization calibration
measurements.  Cen A also lies well above the galactic plane, is well
matched to the size of typical CMB polarimeter beam sizes and is well
studied at many other wavelengths, particularly in the radio.  In this
paper we present the results of observations of Cen A with QUaD, a CMB
polarimeter capable of simultaneously measuring Stokes' $I$, $Q$ and
$U$ parameters with bands at 100 and 150 GHz and angular resolution
$5'.0$ and $3'.5$ respectively.

\section{Instrument summary and observations}
\label{S:observations}

A full description of the QUaD instrument can be found in
\citet{Hinderks2009} (hereafter H09).  QUaD was a $2.6\,$m Cassegrain
telescope using the pre-existing DASI mount at the US south pole
station \citep{Leitch2002}.  QUaD operated from 2005 February to 2007
November, taking astronomical data during the austral autumn through
winter into spring; maintenance and calibrations were performed during
the austral summer.  The receiver comprised 31 pairs of polarization
sensitive bolometers (PSBs), 12 pairs at 100 GHz and 19 pairs at 150
GHz.  At either frequency these were split into two polarization
groups which measure either $Q$ or $U$; each bolometer pair is also
sensitive to Stokes' $I$.

The data presented here were obtained on 2006 June 7 and 2007 October
2, 3 and 4.  The observations were performed in a similar manner to a
standard QUaD observation block as described in \citet{Pryke2009}
(hereafter P09), although the lead-trail scan strategy used in other
QUaD observations is not employed.  Each scan group begins with an
elevation nod during which the telescope is moved up and down by
$1^{\circ}$ in elevation to inject a signal of common amplitude into
each bolometer.  The telescope is then scanned back and forth at
constant elevation over a throw of $3^{\circ}\!\!\!.2$ in sidereal tracking
corrected azimuth; four such scans constitute a single scan block.  For
these observations, the scan rate is $0^{\circ}\!\!\!.066$ per second in
azimuth, or $0^{\circ}\!\!\!.05$ per second on the sky at Cen A's elevation.
After each constant elevation scan block, the elevation is stepped by
1.2 arcmin and the next block is begun.  In all, 101 scan blocks were
performed, equating to a rastered data set mapping $\sim 4^{\circ}
\times 2^{\circ}$ on the sky to a uniform noise level.  These 101
scans required 15.6 hr; adding the standard QUaD calibration set
before and after the scan block (see H09 for a description of such a
set) leads to a total observation time of 17.6 hr per day.

The DASI mount supports rotation of the entire telescope about its
boresight; on a given day this rotation was fixed for the Cen A
observations.  However, the rotation was varied over different days,
which has the benefit of changing the polarization characteristics of
the instrument in a known manner.  The angles chosen were $\{
0^{\circ}, -30^{\circ}, +30^{\circ}, 0^{\circ} \}$ for 2006 June 7 and
2007 October 2, 3, and 4, respectively.

\section{From Time Streams To T Maps}
\label{S:toTmaps}

Unlike standard CMB observations, the QUaD polarization maps of Cen A
require systematic effect corrections derived from the temperature
maps (Section \ref{S:corrections}), so it is necessary to utilize a
two step process where the total intensity maps and then the
polarization maps are constructed.  The Cen A temperature maps
themselves can be constructed directly from the data time series
without any such corrections.

In order to produce maps, the relationship between the measured time
series $d(t)$ and the sky temperatures $T, Q, U$ must be known.
Neglecting noise contributions and the bolometer transfer function, it
can be written as
\begin{multline}
d(t) = g(t) \int d \mathbf{r} \mathcal{B} ( \mathbf{r} -
\mathbf{r}_{b}) \times \\ \left[ T(\mathbf{r}) + \frac{1 - \epsilon}{1
    + \epsilon} \{ Q(\mathbf{r}) \cos 2 \psi + U(\mathbf{r}) \sin 2
  \psi \} \right]
\end{multline}
where $\mathbf{r}$ is the direction of observation and $\psi$ is the
detector's polarization angle (the time dependence of these two
quantities is suppressed for clarity).  The time dependent gain $g$,
cross polar leakage $\epsilon$ and polarized beam shape $\mathcal{B}$
are all quantities which must be determined via calibration
measurements, as must be the zero point of $\psi$.

Initial low-level processing of the time series is performed using the
same algorithm as presented in P09.  The bolometer time constants and
electronic filters are first deconvolved to reverse their effect on
the time series.  These time series are then despiked by removing
those individual scans which exhibit transient impulse events from the
analysis.  The elevation nods are used to determine $g(t)$ for each
scan block by using the changing atmospheric loading to compute the
relative responsivity of each bolometer in a frequency group.  Each
detector in a frequency group is scaled to the group's mean, yielding
relatively calibrated time series for the entire array.

Three more pieces of information are required to create maps from the
rectified and responsivity calibrated bolometer time ordered data
(TOD): the telescope pointing, the PSB angles, and the PSB
efficiencies.  Knowing the pointing of the telescope requires
knowledge of both the sky offset of each of the detectors and the
absolute pointing of the boresight, the derivation of these parameters
is described in detail in H09.  The detector offset angles used in the
map making process are derived from monthly measurements and have an
estimated uncertainty of $~0'.15$; since the signal-to-noise ratio on
Cen A is large the absolute per day pointing solution can be directly
determined from $I$ maps.

The PSB polarization angles and efficiencies were determined using a
chopped thermal source placed behind a polarizing grid on a mast near
the telescope; H09 reviews the measurement and preliminary analysis of
these data.  The cross polar leakage is defined as the ratio of
response to anti-aligned and co-aligned incident polarized light for a
given PSB.  The measured values of $\epsilon$ have a mean of $0.08$
with an rms scatter of $0.015$.

Knowledge of the error on the absolute polarization angle has improved
over that presented in H09 thanks to both an improvement in the
analysis of the calibration data and a new method of measuring this
angle.  The new method of angle measurement relies on the fact that
the polarized response of the detectors is mechanically constrained to
lie along lines of symmetry of the detector rows, any misalignment is
estimated to be random and $<0^{\circ}\!\!\!.5$ per detector.  Measurement
of the polarization angle is performed by scanning rows of detectors
across bright sources at telescope rotation angles where the polarized
response of the detectors matches the orientation of the rows.  As
there are several such symmetries, a number of independent measurements
can be made.  The final error ascribed to this angle measurement
method is $0^{\circ}\!\!\!.15$.

The improved analysis of the calibration source data presented in H09
is based on the observation that the orientation of the polarized
source was not well constrained in the azimuthal plane; azimuthal
misalignments at the source box can result in spurious polarization
measured at the telescope.  We therefore discard all measurements
performed with the calibration source grid in the horizontal
orientation and use only the vertical grid measurements; these are
immune to such alignment errors and have a well measured orientation
with respect to gravity.  The updated analysis which uses only the
grid-vertical measurements has a scatter of $0^{\circ}\!\!\!.2$ over the
measurements.  The absolute polarization angle of the telescope
derived from these measurements is in good agreement with the angle
derived from the row scans.  We therefore use this revised absolute
telescope angle and conservatively ascribe an error of $\pm
0^{\circ}\!\!\!.5$ to its measurement.

To construct maps further processing of the TOD is required.  The PSB
pair time streams are first summed to create total intensity TOD
corresponding to the Stokes' parameter $I$ and differenced to make
instrument frame Stokes' $Q$ or $U$; these must be rotated to an
absolute reference frame for polarization using the known polarization
angle of the telescope (as discussed in Section \ref{S:corrections}).
Drifts in the atmospheric emission and instrument create $1/f$ noise
in the TOD much larger than photon noise on long time scales; to
remove these correlated signals a fifth-order polynomial is fit to
each 48 s half-scan and subtracted from the sum and difference TOD.
This is a higher order than that chosen for the CMB data since we are
not attempting to measure large spatial modes in the Cen A maps.  As
the signal-to-noise ratio on the source is large in a single scan, it
is necessary to mask source flux to prevent biasing the polynomial
fits.  The mask used here comprises a circular region with
$r=0^{\circ}\!\!\!.2$ centered on Cen A; this size was chosen to
preserve structure in the inner lobe region while maximally removing
drifts in the TOD.  The same mask is used for the sum and difference
TOD.  When filtering the sum data to produce $I$ maps, the fit mask is
also augmented by two circular regions centered at R.A.$=\{
13^{\mathrm{h}}\!\!\!.450, 13^{\mathrm{h}}\!\!\!.353 \}$, decl.$= \{
-42^{\circ}\!\!\!.80,-43^{\circ}\!\!\!.70 \}$; these mask Cen A's
moderately bright northern spur region and the elliptical galaxy NGC
5090 from the polynomial filter.

Maps are made by binning the TOD into a grid of pixels weighting by
the inverse variance of each half-scan computed after the polynomial
filter has been applied.  The map pixelization is in R.A.~and
decl.~using square pixels $0^{\circ}\!\!\!.02$ on a side.  The per day
signal-to-noise ratio in the $I$ maps is large, so by comparing
individual day maps it is possible to observe the telescope pointing
solution vary by up to $30$ arcsec day to day.  To correct for this
variation, the per day $I$ maps are made and that day's astrometric
offsets are found by minimizing the sum of the squared pixel
differences in the region of Cen A between that day's map and the map
derived from the first day's observation with its astrometric
calibration manually set to Cen A's known position.  Once determined,
these offsets are then always applied to both sum and difference TOD
during the map making process.

\begin{figure*}[ht]
\centering
\resizebox{0.99\textwidth}{!}{\includegraphics{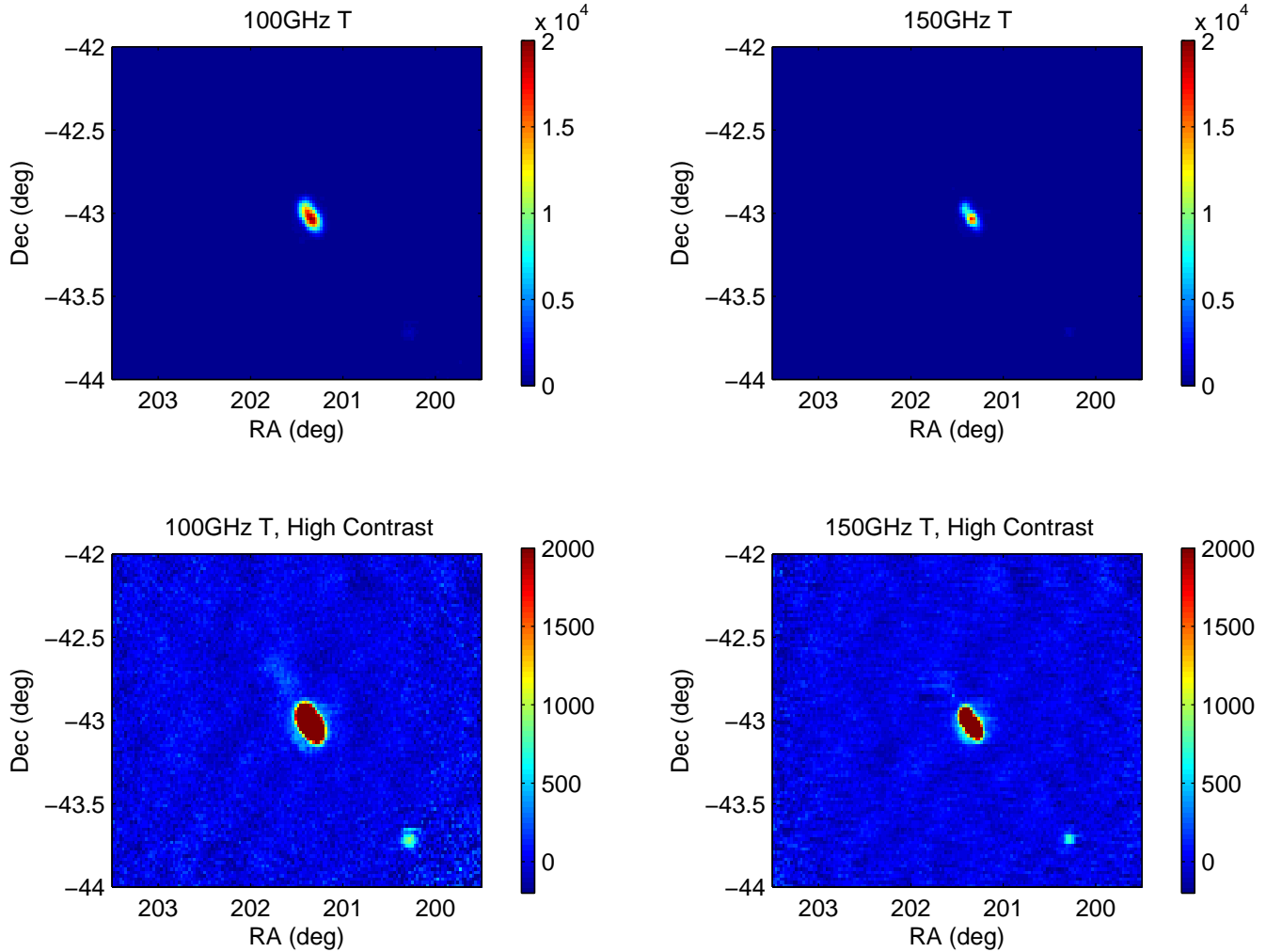}}
\caption{Temperature maps of Centaurus A.  Temperature scale is
  $\mu$K$_{\mathrm{CMB}}$.  \textit{Top:} 100 GHz (left) and 150 GHz
  (right) maps of the Cen A region at low contrast.  The R.A.~and
  decl.~range has been chosen to facilitate comparison with Figure 3
  of \citet{Israel1998} and Figure 1 of \citet{Israel2008}; none of
  the large scale structure visible in those maps at lower frequencies
  is present in the QUaD data.  The structure of the inner lobe region
  comprising two synchrotron emitting convexities in the ICM is
  unresolved at 100 GHz but resolved at 150 GHz.  \textit{Bottom:} 100
  GHz (left) and 150 GHz (right) maps of Cen A at high contrast.  Low
  surface brightness emission is visible to the north east of the
  inner lobe region, as is NCG 5090, an unrelated elliptical galaxy in
  the Cen A field near $\alpha_{\mathrm{J2000}}=200^{\circ}\!\!\!.266,
  \delta_{\mathrm{J2000}}=-43^{\circ}\!\!\!.741$.}
\label{fig:Tmaps}
\end{figure*}

To calibrate to CMB mK we utilize the CMB power spectrum cross
calibration procedure presented in \citet{Brown2009}; the resulting
temperature maps are shown in Figure \ref{fig:Tmaps}.  To calibrate
between $I$ maps in Jy sr$^{-1}$ and $T$ maps in thermodynamic
temperature the relation
\begin{equation}
d I = (dB/dT)_{2.73 \mathrm{K}} d T
\end{equation}
is used.  These values correspond to $\{ 29.2, 58.5 \}\,$Jy map
pixel$^{-1}$ K$^{-1}$ at $\{100, 150 \}$ GHz.

\section{Polarized Instrumental Effect Correction}
\label{S:corrections}

There are a wide range of systematic effects which can mix $I$ to
polarization and $Q$ to/from $U$ in polarimeters.  The QUaD simulator,
discussed in detail in P09, is an extremely detailed model of the
instrument and various sources of noise whose primary purpose is to
model time streams for use in {\sc master}-style CMB power spectra
analysis \citep{Hivon2002}.  As the simulator is built to model the
instrument accurately it can be used to measure and correct for
instrumental systematic effects which are known to exist in QUaD data,
whatever the astronomical source under study.  In order to produce
clean polarization maps, it is necessary to simulate and correct for
these polarization mixing effects.

\subsection{Polarization leakage correction}
\label{sS:polleakage}

Although the detector offset angles show no evidence for variation
over time or between bolometer pairs sharing a feed horn, they do
exhibit repeatable offsets between the two halves of each detector
pair with an rms magnitude of $~0'.1$ over the array.  When the sum
and difference maps are constructed, the mean detector offset of a
pair is used; unfortunately, the difference between this mean and the
actual pointing of a detector can mix $I \rightarrow Q, U$.  This
polarization leakage can be quantified using simulations of the
observations.

To measure polarization leakage, an input map is generated from the
coadded $I$ maps at each frequency shown in Figure \ref{fig:Tmaps}.
Hereafter, we denote the ``polynomial filter mask region'' to be that
area in the map corresponding to the regions masked in the time stream
polynomial fit discussed in Section \ref{S:toTmaps}, i.e.~the
$r < 0^{\circ}\!\!\!.2$ area centered on the source.  Regions outside of the
polynomial filter mask region are set to zero, such that only the
central area contains non-zero flux.  Each detector's pointing time
stream is constructed using the individual measured offsets, which
vary between detectors in a pair.  These time series are used to
sample a new, simulated time stream from the measured $I$ maps.  As no
noise is added to the TOD and the polarization maps are set to zero,
this procedure generates a simulation of the pure $I$ component of the
Cen A observation for each detector.  The PSB pairs are then
differenced in the standard way; if the pointing between pairs was
perfect, this procedure would cancel the structure appearing in the
constructed $I$ maps perfectly.  However, as the pair pointing is not
identical, this procedure produces residuals reflecting the leakage of
$I$ to polarization in the actual observation.  These leakage time
streams are subtracted from their corresponding pairs in the data TOD,
which corrects for the leakage effects at the time sample level.

\begin{figure*}[ht]
\centering
\resizebox{0.99\textwidth}{!}{\includegraphics{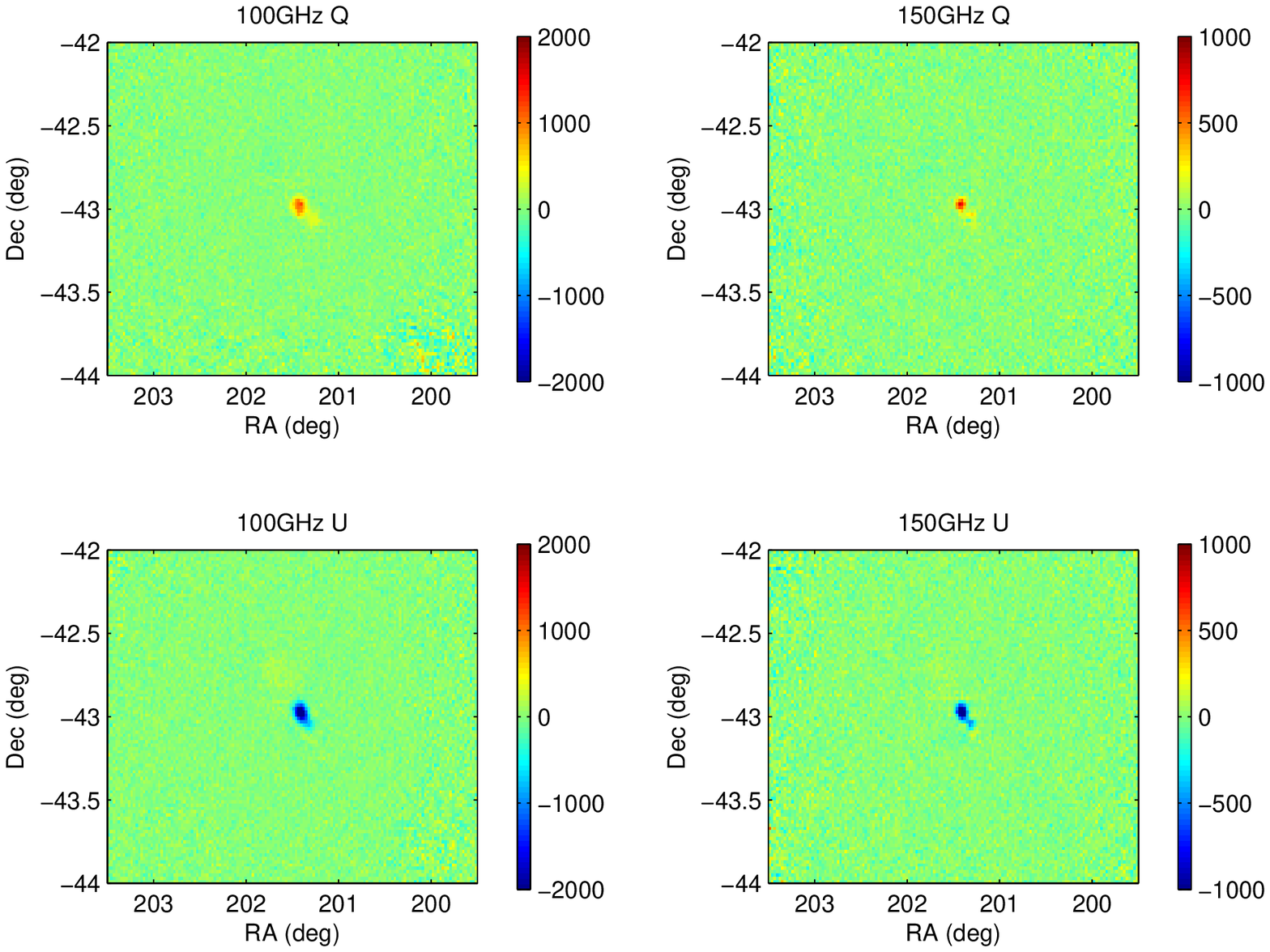}}
\caption{Polarization leakage corrected maps of Centaurus A.
  Temperature scale is $\mu$K$_{\mathrm{CMB}}$.  \textit{Top:} 100 GHz
  (left) and 150 GHz (right) maps of the Cen A region for Stokes' $Q$.
  \textit{Bottom:} 100 GHz (left) and 150 GHz (right) maps of the Cen
  A region for Stokes' $U$.  The polarization of the northern inner
  lobe is about 20 \% of $I$ at either frequency.  The polarization
  fraction in the southern lobe is much smaller.}
\label{fig:Pmaps}
\end{figure*}

These polarization leakage corrected time streams are mapped onto the
same R.A.~and decl.~grid as the $I$ data.  For difference time stream
data, the product of the data and the sine and cosine of the detector
angles as projected on the sky must be accumulated for each pixel to
weight the different detector pairs according to their polarization
sensitivities; this procedure rotates instrument frame $Q$ and $U$ to
absolute $Q$ and $U$.  The set of $2 \times 2$ matrices comprising
products of these sines, cosines and difference time streams must be
inverted for each pixel; these are then accumulated into the grid to
produce absolutely referenced $Q$ and $U$ maps.  In this work we
follow the IAU polarization convention with positive $Q$ running
north--south and positive $U$ running northeast--southwest
\citep{Hamaker1996}.  Figure \ref{fig:Pmaps}\ shows the CMB
temperature calibrated $Q$ and $U$ maps at 100 and 150 GHz.  These,
combined with the $I$ maps, are the fundamental output of the analysis
pipeline.

These maps can be used to characterize the polarization of Cen A per
QUaD beam or per map pixel as desired (see Section \ref{S:results}).

\subsection{Polynomial filter correction}
\label{sS:filtercorrection}

A second correction which needs to be applied to these data is due to
the time series polynomial filter.  The filter mask discussed in
Section \ref{S:toTmaps}\ is constructed as a compromise between
maximizing the mask size to exclude as much source flux as possible
from the polynomial filter fit and the need to minimize the mask so
that individual scans are well constrained when the polynomial
function is determined from the data.  Although the mask effectively
excludes the bright source flux during the polynomial filter fit,
there always remains some small fraction of the source flux outside
the mask for the scan length used in these measurements.  This flux
has the effect of biasing the filter determined from the data; this
can be quantified using the QUaD simulator.

In order to produce realistic simulations of these observations, it is
necessary to create an accurate source map at each polarization and
frequency.  Such a map needs to have high resolution so that the
effect of the QUaD beams can be accurately simulated, and must have
large enough signal-to-noise ratio on all components of the source
that the input map is injecting negligible error into the simulation.
In the case of the CMB highly accurate simulations of the background
sky based on known physical mechanisms at the epoch of recombination
exist; unfortunately, the same is not true of Cen A.  Such a map of
the Cen A field at QUaD's frequencies can be produced either by
scaling a map from a different frequency by the relevant spectral
index of the source, or by using the QUaD maps themselves.  The former
method is difficult: maps of the Cen A region at radio frequencies
tend to either have much higher resolution but are not sensitive to
extended emission, or much lower resolution which would not allow us
to resolve the source.  We therefore have utilized the latter
approach.

In order to increase the angular resolution of the Cen A maps, an
algorithm which deconvolves the effective beam from the observed maps
is necessary.  We have chosen the Richardson-Lucy deconvolution (RLD)
algorithm here as it can efficiently deconvolve a known kernel from
noisy data (\citealt{Richardson1972}; \citealt{Lucy1974}).  The RLD
algorithm employs Bayes' theorem to iteratively reconstruct the
maximum likelihood background map given the beam kernel and the
observed map with which it has been convolved.  For the QUaD
measurements of Cen A, the input maps are the observed 4 day coadded
maps shown in Figures \ref{fig:Tmaps}\ and \ref{fig:Pmaps}.

\begin{figure}[ht]
\centering
\resizebox{0.48\textwidth}{!}{\includegraphics{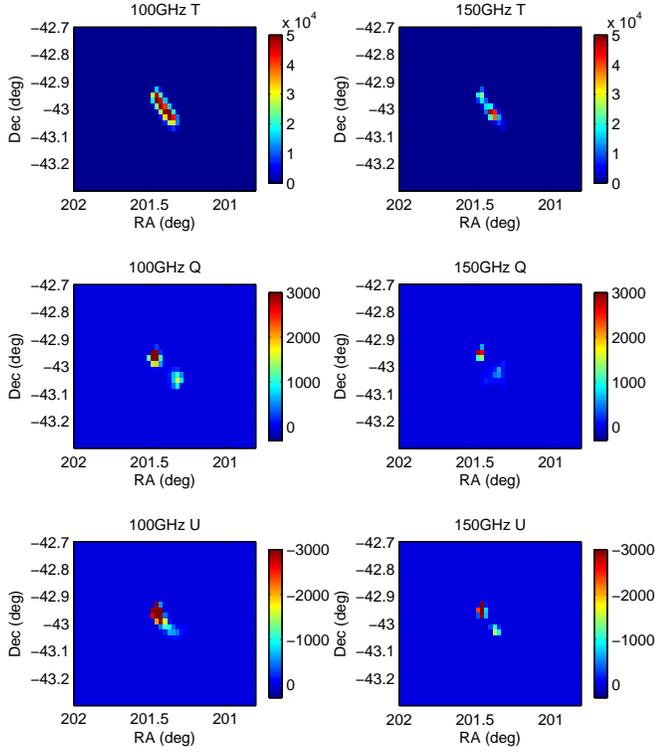}}
\caption{RLD maps of Centaurus A.  The deconvolution kernel is a
  symmetric Gaussian with an FWHM matching the measured QUaD beam.
  Rows correspond to $T$ (top), $Q$ (middle), and $U$ (bottom) while
  columns are for $100$ GHz (left) and $150$ GHz (right).  Temperature
  scale is $\mu$K$_{\mathrm{CMB}}$; the scaling on the $U$ maps has
  been reversed to allow comparison with the $Q$ maps.  Note that the
  deconvolved maps resolve the source into two distinct lobes with
  little or no polarized emission arising from the nuclear region
  between them, particularly at 150 GHz.}
\label{fig:RLDmaps}
\end{figure}

The effective QUaD beam for these observations has been calculated
using noiseless simulations whose input is a delta function.  These
computed beams are well matched to a Gaussian to a level below $-15
\,$dB, so for computational ease we choose a RLD beam kernel to be a
symmetric Gaussian with FWHM matching the QUaD beams at either
frequency (H09).  Applying the RLD algorithm to the 4 day coadded Cen
A images yields the maps shown in Figure \ref{fig:RLDmaps}.  The
deconvolution is performed for each polarization state at both
frequencies; in the case of $Q$ and $U$ the polarization leakage
corrected maps are used in the deconvolution so that polarization
leakage need not be simulated.

The polynomial filter correction is determined by performing
simulations similar to those discussed in the previous section.  The
RLD input maps are convolved with the ideal instrument response
function at each frequency and TOD are constructed for each detector
using the known telescope pointing.  These TOD are then summed and
differenced pair-wise, and a fifth order polynomial is fit and
subtracted from each pair's time streams.  These filtered TOD are then
binned in the usual way to produce fully corrected polarization maps; 
these constitute QUaD's final processed maps of Cen A.

\section{Results}
\label{S:results}

\begin{figure*}[ht]
\centering
\resizebox{0.98\textwidth}{!}{\includegraphics{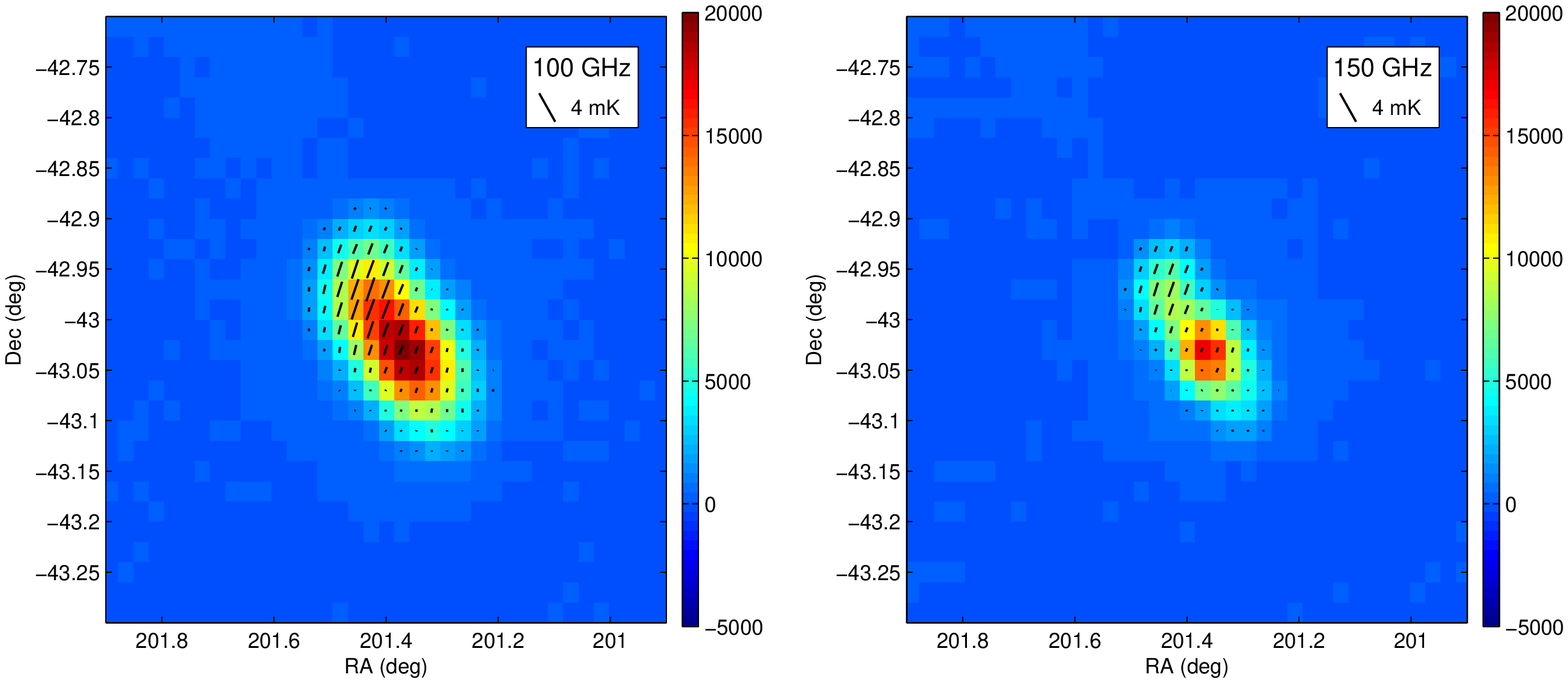}}
\caption{Final $T$ maps with fully processed polarization vectors over
  plotted.  The left- and right-hand panels show the 100 and 150 GHz
  maps, respectively.  Polarization vectors are plotted for those map
  pixels with $T > 1000 \, \mu$K.  The color scale is in $\mu$K and
  the length of a $4\,$mK pure $+U$ vector is indicated.}
\label{fig:goodmaps}
\end{figure*}

Figure \ref{fig:goodmaps} shows the 4 day co-added maps after the
systematic corrections have been applied.  Using the RLD algorithm,
these can be deconvolved with the nominal QUaD beam to remove the
effect of the instrument transfer function to yield the maps similar
to those shown in Figure \ref{fig:RLDmaps}; these maps allow
simulations of this source in $I, Q$ and $U$ for any instrument with
beam size similar to QUaD's or larger\footnote{Both systematic
  corrected as-measured and RLD maps, including estimates of the noise
  in each map pixel, will be publicly available on the web at
  \tt{http://find.uchicago.edu/quad/quad\_CenA/}.}.

A quantity which encapsulates a great deal about these polarization
maps is the total $Q$ and $U$ in an aperture.  Here we employ a circular
aperture with $r=0^{\circ}\!\!\!.2$ centered on Cen A's position; this
matches the source mask used during the time stream polynomial
filtering.  The total polarized temperatures for the final 4 day
co-added maps are given by
\begin{equation}
\label{eq:scriptyP}
\mathcal{P} = \{ \sum_{r} Q(\alpha, \delta), \sum_{r} U(\alpha, \delta) \}.  
\end{equation}
The values of the aperture sums $\mathcal{P}$ measured from the QUaD
maps are listed in Table \ref{tab:results}.  The error estimate from
the map, denoted $\sigma_{\mathrm{map}}(\mathcal{P})$, is computed by
calculating the rms of pixels in the same circular aperture offset
from the source by $\Delta \mathrm{R.A.} = 0^{\circ}\!\!\!.5$ and dividing by
the square root of the number of map pixels in the sum region.  The
error estimates $\sigma_{\mathrm{map}}(\mathcal{P})$ are computed
separately for $Q$ and $U$ at both frequencies.

\begin{table}[ht]
\centering
\caption{Polarized Flux of Cen A from
  Corrected QUaD $Q$ and $U$ Maps. }
\begin{tabular}{l|cc}

 & 100 GHz & 150 GHz \\ \hline
$\sum_{r} Q$ & 1.00 Jy & 0.80 Jy \\ 
$\sigma_{\mathrm{map}}(\sum Q)$ & 0.06 Jy & 0.05 Jy \\
$\sigma_{\mathrm{sim}}(\sum Q)$ & 0.06 Jy & 0.06 Jy \\ \hline
$\sum_{r} U$ & -1.72 Jy & -1.40 Jy \\
$\sigma_{\mathrm{map}}(\sum U)$ & 0.08 Jy & 0.05 Jy \\
$\sigma_{\mathrm{sim}}(\sum U)$ & 0.06 Jy & 0.07 Jy \\  \hline
$\theta$ & $-30^{\circ}\!\!\!.0$ & $-29^{\circ}\!\!\!.1$ \\
$\sigma (\theta)$ & $0^{\circ}\!\!\!.9$ & $1^{\circ}\!\!\!.2$ \\
\hline
\end{tabular}
\label{tab:results}
\end{table}

An independent measurement of the error in $\mathcal{P}$ can be
provided by the QUaD simulator, which includes an extremely accurate
model of the noise in QUaD data.  As in previous simulations, the RLD
input map is convolved with the individual detector's beams and
sampled with the telescope pointing.  Based on each detector's
statistical properties over a noise measurement block, noise is
generated and injected into the scan TOD.  P09 details the noise model
construction in detail; it is useful to note here that the noise
estimate for each detector is constructed from statistical blocks of
that detector's actual time series 5 scans long.  These time streams
are then fifth order polynomial filtered and binned into $I, Q, U$
maps as with the observed data.  This simulation process is repeated
$N$ times to obtain $N$ random realizations of the measurement.  The
scatter of $\mathcal{P}$ in these $N$ realizations yields the
variation in the possible measurement outcomes due to the random noise
in the data.

\begin{figure}[ht]
\centering
\resizebox{0.48\textwidth}{!}{\includegraphics{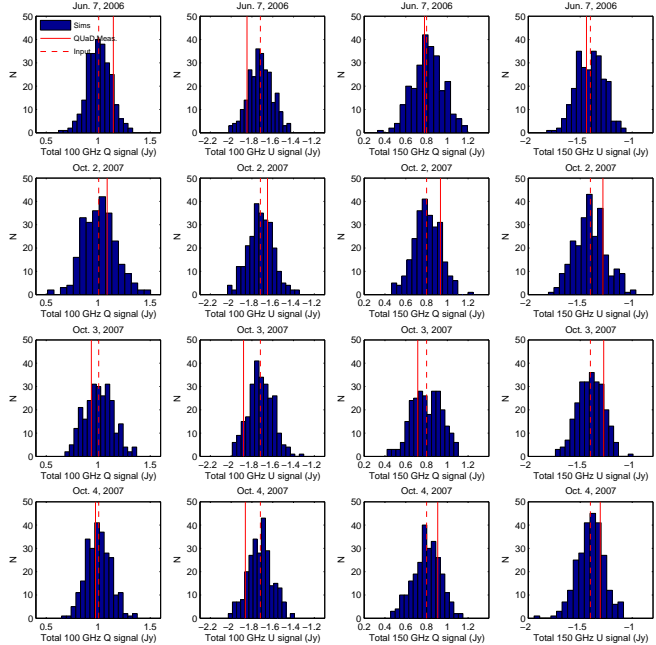}}
\caption{Results of $256$ simulations of the per day Cen A
  observations for $Q$ and $U$ at 100 and 150 GHz.  Days run row-wise
  and polarization and frequencies run column-wise.  The simulations
  use knowledge of the individual detectors' beams, the full QUaD
  noise model, and share the same analysis pipeline as the actual
  data.  The individual realizations of $\mathcal{P}$ at both
  frequencies are shown in the histograms; also shown are the input
  $\mathcal{P}$ (dotted lines) and the $\mathcal{P}$ measured from the
  data (solid lines) for each day, frequency, and polarization.  This
  result shows that there is no bias in the simulations, that the
  measured data are compatible with random draws from such
  realizations, and that the noise in $\mathcal{P}$ can be estimated
  by the scatter in the simulation realizations.}
\label{fig:perdayhists}
\end{figure}

Figure \ref{fig:perdayhists}\ shows the results of $256$ such
realizations for each day of observation.  The standard deviations of these
distributions are consistent with the aperture rms of the individual
day maps, showing that the noise model is in good agreement with the
variance in the data.  Also plotted in this figure are the corrected
per day $\mathcal{P}$ Cen A measurements; the hypothesis that these
measurements are drawn from the same distribution as the simulated
sample can be checked using Student's $t$-test.  In the sample of 16
different permutations of frequency, $Q$, $U$ and day, the null
hypothesis is never rejected.  This result is consistent with the null
hypothesis for the overall set, and shows both that we have not
detected a bias in the individual days' data and that the variance in
the simulations is a realistic model for the variance in the real
data.  The random error estimate on $\mathcal{P}$ derived from the
scatter of the simulation results is listed in Table \ref{tab:results}
as $\sigma_{\mathrm{sim}}(\mathcal{P})$.

The overall polarization angle of the source can be computed using
\begin{equation}
\theta = \frac{1}{2} \arctan \left( \sum U / \sum Q \right).
\label{eq:polangle}
\end{equation}
The values obtained in the QUaD measurement are listed in Table
\ref{tab:results}, including the random error $\sigma (\theta)$.  The
released data maps and equation \ref{eq:polangle}\ can be applied per
map pixel to determine the polarization angle variation over the
source.

\section{Systematic issues}
\label{S:systematics}

There are a number of systematic effects which can add error to the
measurement of $\mathcal{P}$ and $\theta$.  These broadly group into
either astrophysical effects or instrumental effects; these are
discussed below.  

\subsection{Background Structure}
\label{sS:background}

The background structure in the Cen A field can potentially cause
significant systematic error in this measurement.  A number of
measurements have mapped out Cen A's $8^{\circ} \! \times 4^{\circ}$
radio emitting outer lobes at various frequencies, most recently at
high radio frequencies with \textit{WMAP} \citep{Wright2009}.
\citet{Israel2008} show detailed \textit{WMAP} maps of Cen A at 23,
33, 41, 61 and $94 \,$GHz; the outer lobe emission is clearly visible
at the lower frequencies but is not significantly detected at $94
\,$GHz.  This result is evidence that at the QUaD frequencies Cen A's
outer lobe synchrotron emission is much below the brightness of the
inner lobe region, but this result should be checked.

\begin{figure}[ht]
\centering
\resizebox{0.48\textwidth}{!}{\includegraphics{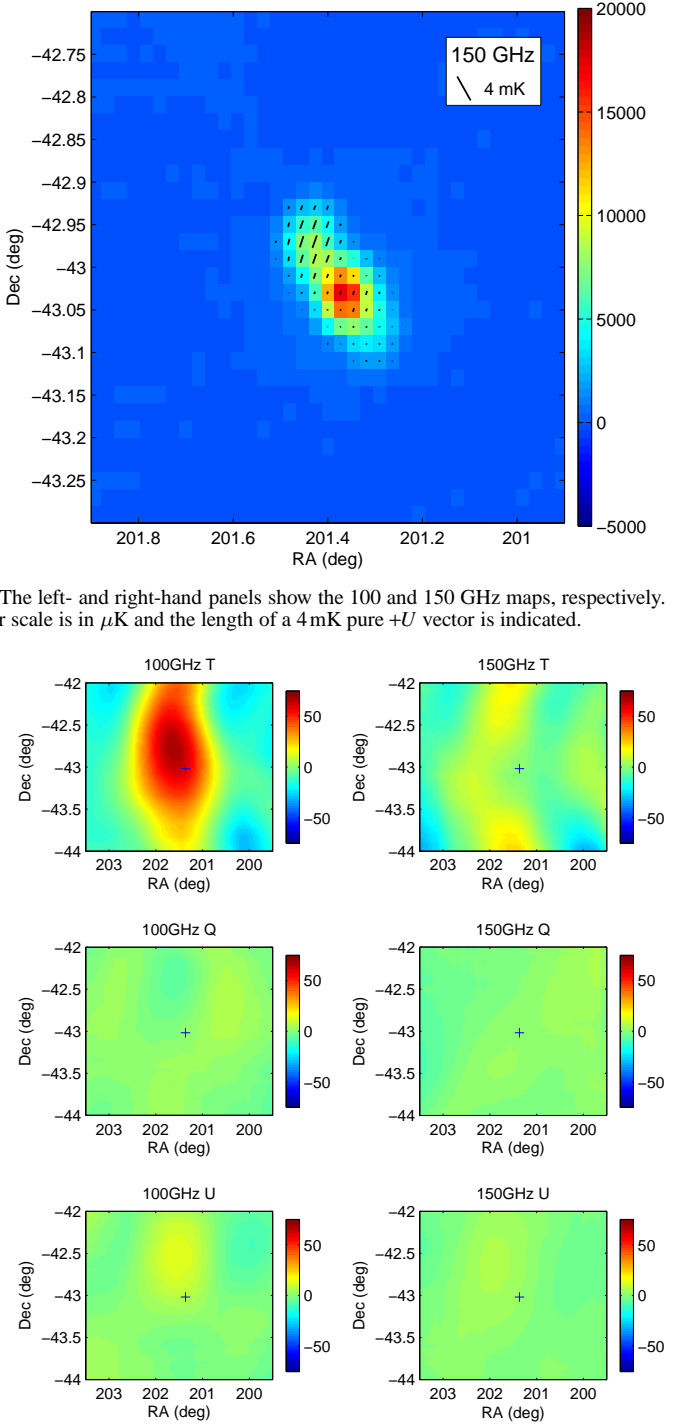}}
\caption{Background-filtered maps of the Cen A region.  The left-hand
  column shows 100 GHz maps for $T$ (top), $Q$ (middle), and $U$
  (bottom); the right-hand column is the same for the 150 GHz maps.
  Color scale is in $\mu$K and the cross in each map shows the nominal
  position of Cen A.  The 100 GHz $T$ map is the only one exhibiting
  evidence of background structure; there is no evidence for
  background contamination in the polarization maps.}
\label{fig:background}
\end{figure}

To perform this assessment, maps are first made in the standard way,
except a first order polynomial filter is used to remove the
instrumental drifts from the time series of each detector; unlike the
standard fifth-order polynomial used in the normal map making process,
this filter preserves structures larger than $1^{\circ}$ on the sky.
The central $r < 0^{\circ}\!\!\!.2$ region is again masked during the
polynomial fitting procedure to avoid bias.  These data are binned into
maps, and then the central polynomial mask region is replaced with the
mean of the pixels in the annulus $0^{\circ}\!\!\!.2 < r < 0^{\circ}\!\!\!.3$ just
beyond the mask region.  These new maps are then convolved with a
Gaussian kernel with an FWHM of $1^{\circ}$; this smooths out the small
features whilst retaining the largest structures in the map.  Figure
\ref{fig:background} shows the resulting maps for $T$, $Q$ and $U$.
Of these, only the 100 GHz $T$ map shows evidence for background
structure on scales $\sim 1^{\circ}$.  The total flux in the region
where $T > 20 \mu$K in this background-filtered 100 GHz map is 10\% of
the total flux in the inner lobe region at 100 GHz.  A number of
different kernel sizes from $0^{\circ}\!\!\!.2$ to $2^{\circ}$ have been
applied in this procedure and the total flux in the background region
does not change appreciably under different kernel widths.  Table
\ref{tab:systematics}\ lists the estimated uncertainty in
$\mathcal{P}$ and $\theta$ caused by the background structure; for
each polarization and frequency this quantity is calculated by summing
the background signal in the region where $T_{100 \mathrm{GHz}} > 20
\mu$K and multiplying the total by the ratio of the area of the $r <
0^{\circ}\!\!\!.2$ region to the area of the summed region.

\subsection{Source Variability}
\label{sS:variability}

Another potential source of systematic error is temporal flux
variability in Cen A itself.  Although the inner lobe region is too
large for its emission to change on human time scales, the nucleus of
the source is known to be variable in $I$.  Further, no measurements
of this source's variability in polarization around 100 GHz exist.

\begin{figure}[ht]
\centering
\resizebox{0.48\textwidth}{!}{\includegraphics{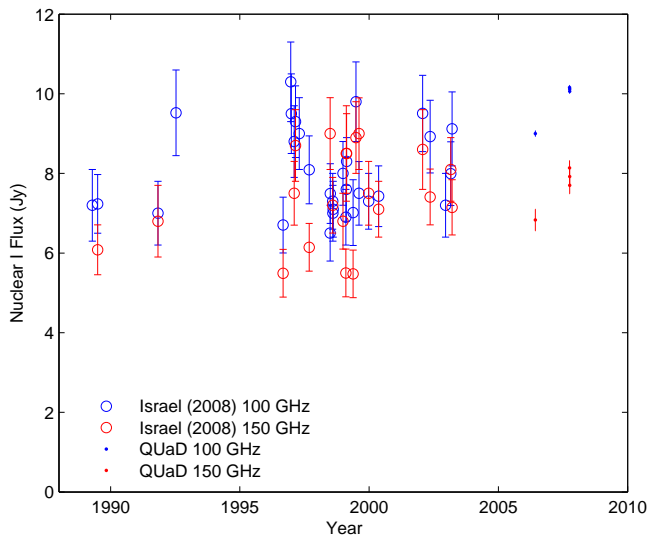}}
\caption{Variability in Centaurus A's nucleus at 100 and 150 GHz for
  Stokes' $I$.  The open circles show data presented in
  \citet{Israel2008} for the period 1988--2004 obtained using a
  heterodyne system on the SEST telescope, while the dots show the
  QUaD data presented here.  These data are compatible with the
  well-known $\sim 30 \,$\% peak to peak radio variation of this
  source at these wavelengths.}
\label{fig:Tvar}
\end{figure}

To measure the $I, Q$ and $U$ variability in Cen A, the nominal QUaD
beam shape centered on the Cen A nucleus is fit to the final processed
beam convolved maps at each frequency.  As the size of Cen A's nucleus
is much smaller than a QUaD beam, this fit yields the flux in the
nuclear region, and subtracting it from the map yields the total flux
in the inner lobe region for each frequency and polarization.  The
error in the resulting nuclear flux is calculated by measuring the rms
deviations of the pixels in a number of circular apertures which have
the same angular extent as a QUaD beam.  The set of aperture rms
statistics is then averaged to yield the noise in the measurement of
the flux in a single aperture.  Figure \ref{fig:Tvar} shows the $I$
measured by QUaD in Cen A's nuclear region for each day of observation
at 100 and 150 GHz.  At both QUaD's bands, the $I$ flux from Cen A's
nucleus is $\sim 35$\% of the total flux from the inner lobe and
nuclear regions; this implies that the temporal variability expected
in the QUaD maps is $\lesssim 10$\% in $I$.  Also plotted are the
nuclear fluxes measured by \citet{Israel2008}.  These measurements
were performed with a heterodyne system on SEST at several frequencies
between 80 and 300 GHz: in order to create data comparable to the QUaD
set, we compute the QUaD bandpass-weighted average of the SEST points.

The \citet{Israel2008} points clearly exhibit the $\approx 30 \,$\%
peak to peak variation in $I$ well known from measurements at lower
radio frequencies.  The QUaD measurements also exhibit these
fluctuations, although essentially only at two independent times as
the 2007 measurements were performed on 3 consecutive days.  These
data do place a limit on the short term (day time scale) $I$
variability of the Cen A nucleus of $1 \,$\% at 100 GHz and $6 \,$\%
at 150 GHz.  The $I$ flux of only the inner lobes in the standard $r <
0^{\circ}\!\!\!.2$ aperture is measured to be $14.2 \pm 0.3 \,$Jy at $100
\,$GHz and $16.6 \pm 1.1 \,$Jy at $150 \,$GHz.

\begin{figure}[ht]
\centering
\resizebox{0.48\textwidth}{!}{\includegraphics{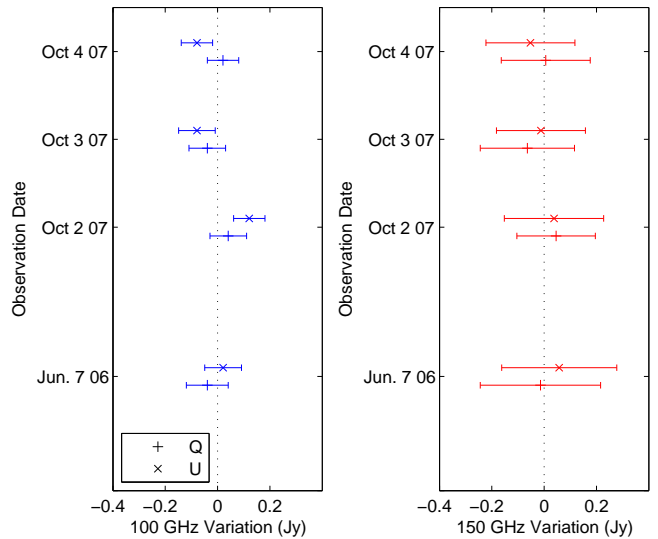}}
\caption{Time variability of Cen A in total $Q$ and $U$ at 100 and 150
  GHz (left and right, respectively).  Symbol ``$+$'' shows $Q$ and
  symbol ``$\times$'' shows $U$ for each of the four days of
  measurement; flux is given in Jy.  The mean of each frequency and
  polarization group has been subtracted to isolate the variable
  component of the flux.  As these points are consistent with no time
  variation, half of their peak to peak scatter is used as the error
  associated with variability in the measurement of Cen A's
  polarization properties.}
\label{fig:Pvar}
\end{figure}

Cen A's nuclear flux in $Q$ and $U$ is computed in the same way as for
$I$; Figure \ref{fig:Pvar}\ shows the resulting $Q$ and $U$ for the 4
days of observation.  Comparing to the model that the flux of these 16
measurements is constant, the obtained reduced $\chi^{2}=0.64$; based
on the QUaD data, there is no evidence of polarized temporal
variability in Cen A.  An upper limit on the possible variation can be
obtained by computing the mean absolute deviation of these individual
day data from their mode; the limits are listed in Table
\ref{tab:systematics}.  These limits are significantly smaller than
the estimated 30\% variation expected from the $I$ variability.
Inspection of the RLD maps in Figure \ref{fig:RLDmaps}\ shows that
very little of the polarized emission is originating from the nuclear
region itself.  This is evidence that this source's polarization
stability arises because the polarized millimeter-wave flux is
predominantly being emitted by the inner lobe region rather than the
nucleus.

\subsection{Instrumental Parameters}
\label{sS:instrumental}

There are a number of instrumental parameters used in the QUaD
analysis pipeline which can cause systematic errors; a review of these
is given in P09.  Following that work, for these Cen A data the
possible systematic errors caused by misestimates of these 
parameters are investigated using the QUaD simulator.

The absolute calibration uncertainty from detector output to
$T_{\mathrm{CMB}}$ is estimated to be $3.4$\% \citep{Brown2009}.  As
this uncertainty is completely correlated between $I$, $Q$ and $U$ it
does not affect polarization angle or fraction, only absolute
polarization temperature.  For this reason it is omitted from Table
\ref{tab:systematics}, though formally it should be included in a
complete accounting of the uncertainty in Cen A's $Q$ and $U$.

As discussed in Section \ref{sS:polleakage}, the detector pair beam
offsets cause leakage of $I$ to polarization.  Although we have
accounted for this effect, the measurement of the offsets is still
subject to error.  To measure the effect of a random misestimation of
the detector offsets, we have performed simulations where the measured
detector offsets are randomized by the rms centroid uncertainty of
$0'.15$. Though this leads to no change in the mean value of
$\mathcal{P}$, it increases the noise in the measurement by a few
tenths of mK.  We adopt the simulated values as a conservative
estimate of the error due to misestimation of the detector offsets in
Table \ref{tab:systematics}.
\begin{table*}[ht]
\centering
\caption{Cen A measurement uncertainty budget.}
\begin{tabular}{l|cccccc}
Error on parameter & $\sum Q_{100 \mathrm{GHz}}$ & $\sum U_{100 \mathrm{GHz}}$ & 
$\sum Q_{150 \mathrm{GHz}}$ & $\sum U_{150 \mathrm{GHz}}$ & 
$\theta_{100 \mathrm{GHz}}$ & $\theta_{150 \mathrm{GHz}}$ \\ \hline

% sim 11600 - 9'' random
Beam pair offsets & $9$ mJy & $13$ mJy & 
$10$ mJy & $18$ mJy & $0^{\circ}\!\!\!.15$ & $0^{\circ}\!\!\!.22$ \\

% sim 12000 - 2.5% scatter in width
Differential beam shape & $4$ mJy & $5$ mJy & 
$6$ mJy & $6$ mJy & $0^{\circ}\!\!\!.06$ & $0^{\circ}\!\!\!.11$ \\

% sim 13300 - 1 degree random scatter in angle
Relative pair polarization angle & $9$ mJy & $6$ mJy & 
$5$ mJy & $4$ mJy & $0^{\circ}\!\!\!.12$ & $0^{\circ}\!\!\!.09$ \\

Background structure & $1$ mJy & $0$ mJy 
& $4$ mJy & $32$ mJy & $0^{\circ}\!\!\!.01$ & $0^{\circ}\!\!\!.29$ \\ 

Source variability & $14$ mJy & $31$ mJy & 
$56$ mJy & $68$ mJy & $0^{\circ}\!\!\!.28$ & $1^{\circ}\!\!\!.05$ \\

Calibration angle & $35$ mJy & $35$ mJy &
$28$ mJy & $28$ mJy & $0^{\circ}\!\!\!.50$ & $0^{\circ}\!\!\!.50$ \\ 

Total systematic  & $39$ mJy & $49$ mJy & 
$64$ mJy & $83$ mJy & $0^{\circ}\!\!\!.61$ & $1^{\circ}\!\!\!.23$ \\ \hline

Random & $65$ mJy & $57$ mJy &
$64$ mJy & $70$ mJy & $0^{\circ}\!\!\!.91$ & $1^{\circ}\!\!\!.17$ \\ \hline

Total error & $76$ mJy & $75$ mJy & 
$91$ mJy & $109$ mJy & $1^{\circ}\!\!\!.09$ & $1^{\circ}\!\!\!.69$ \\ \hline

\end{tabular}
\label{tab:systematics}
\end{table*}

Although the polarized beam shape $\mathcal{B}$ has been measured to
be symmetric between bolometers sharing a feed, the error of this
measurement is approximately $2.5$\% of the beam width.  The
uncertainty in the individual detector beam widths could cause a
systematic error in this analysis.  As there is no evidence for a
systematic shift in the beam widths, to quantify this error we have
performed simulations where the pair beam widths are scattered about
their known values using a Gaussian randomization with $1 \sigma =
2.5$\%.  In addition, the relative detector pair polarization angle
uncertainty is also a potential source of polarization error.
Simulations are performed assuming a $1 \sigma = 1^{\circ}$ Gaussian
scatter on the measured detector pair relative angles, which is the
estimated per pair uncertainty on $\psi$ (see H09).  The results of
both of these simulation sets are listed in Table
\ref{tab:systematics}.

As discussed in P09, we estimate the polarization leakage of our
detectors to be $\epsilon = 0.08 \pm 0.015$.  Random errors in
$\epsilon$ will average down, while a systematic change merely shifts
the absolute calibration by a factor of $\sim 2 \sigma_{\epsilon}$.
As the uncertainty on $\epsilon$ is $< 0.02$ this is sub-dominant to
the uncertainty on the overall absolute calibration.  Sidelobe pickup
of polarized emission from the ground may also be a concern
\citep{Brown2009}.  However, both use of the fifth-order polynomial
filter and the small angular extent of the source suppresses the
magnitude of this effect to the $< 10 \, \mu$K level, so it is not a
significant source of systematic error in this measurement.

Table \ref{tab:systematics}\ summarizes the error budget for this
measurement including all important systematic errors.  The total
systematic and total errors have been calculated using the square root
of the quadrature sum of each of the individual errors.  

\section{Conclusion}

We have presented measurements of the $4^{\circ} \! \times 2^{\circ}$
region centered on the radio source Centaurus A with QUaD, a
millimeter-wave polarimeter whose absolute polarization angle is known
to $\pm 0^{\circ}\!\!\!.5$.  Systematic errors from astronomical sources
such as the Cen A field's background structure and temporal variability
from the source's nuclear region have been estimated.  Simulations have
been performed to assess the effect of misestimation of the
instrumental parameters on the final measurement.  After correcting
for known instrumental effects, we find that the total $(Q, U)$ of the
inner lobe region is $(1.00 \pm 0.08, -1.72 \pm 0.08) \,$Jy at 100 GHz
and $(0.80 \pm 0.09, -1.40 \pm 0.11) \,$Jy at 150 GHz, leading to
polarization angles of $-30^{\circ}\!\!\!.0 \pm 1^{\circ}\!\!\!.1$ and
$-29^{\circ}\!\!\!.1 \pm 1^{\circ}\!\!\!.7$.

Future millimeter polarimeters, including CMB experiments, will
require a well characterized astronomical source against which to
compare their instruments' laboratory determined polarization
properties.  Cen A represents one of the best candidates for a stable,
compact, highly polarized source, and the measurements presented here
will allow calibration of such instruments to an uncertainty of $\sim
1^{\circ}$.

\section*{Acknowledgements} 

QUaD is funded by the National Science Foundation in the USA, through
grants ANT-0338138, ANT-0338335 and ANT-0338238, by the Science and
Technology Facilities Council (STFC) in the UK and by the Science
Foundation Ireland.  The BOOMERanG collaboration kindly allowed the
use of their CMB maps for our calibration purposes.  M.Z.~acknowledges
support from a NASA Postdoctoral Fellowship.  P.G.C.~acknowledges
funding from the Portuguese FCT. S.E.C.~acknowledges support from a
Stanford Terman Fellowship. J.R.H.~acknowledges the support of an NSF
Graduate Research Fellowship, a Stanford Graduate Fellowship and a
NASA Postdoctoral Fellowship. Y.M.~acknowledges support from a SUPA
Prize studentship. C.P.~acknowledges partial support from the Kavli
Institute for Cosmological Physics through the grant NSF PHY-0114422.
E.Y.W.~acknowledges receipt of an NDSEG fellowship.

\bibliography{ms}

\begin{thebibliography}{12}
\expandafter\ifx\csname natexlab\endcsname\relax\def\natexlab#1{#1}\fi

\bibitem[{{Barkats} {et~al.}(2005){Barkats}, {Bischoff}, {Farese}, {Gaier},
  {Gundersen}, {Hedman}, {Hyatt}, {McMahon}, {Samtleben}, {Staggs},
  {Stefanescu}, {Vanderlinde}, \& {Winstein}}]{Barkats2005}
{Barkats}, D., {et~al.} 2005, \apjs, 159, 1

\bibitem[{{Brown} {et~al.}(2009){Brown}, {Ade}, {Bock}, {Bowden}, {Cahill},
  {Castro}, {Church}, {Culverhouse}, {Friedman}, {Ganga}, {Gear}, {Gupta},
  {Hinderks}, {Kovac}, {Lange}, {Leitch}, {Melhuish}, {Memari}, {Murphy},
  {Orlando}, {O'Sullivan}, {Piccirillo}, {Pryke}, {Rajguru}, {Rusholme},
  {Schwarz}, {Taylor}, {Thompson}, {Turner}, {Wu}, {Zemcov}, \& {The QUa D
  collaboration}}]{Brown2009}
{Brown}, M.~L., {et~al.} 2009, \apj, 705, 978

\bibitem[{{Hamaker} \& {Bregman}(1996)}]{Hamaker1996}
{Hamaker}, J.~P., \& {Bregman}, J.~D. 1996, \aaps, 117, 161

\bibitem[{{Hinderks} {et~al.}(2009){Hinderks}, {Ade}, {Bock}, {Bowden},
  {Brown}, {Cahill}, {Carlstrom}, {Castro}, {Church}, {Culverhouse},
  {Friedman}, {Ganga}, {Gear}, {Gupta}, {Harris}, {Haynes}, {Keating}, {Kovac},
  {Kirby}, {Lange}, {Leitch}, {Mallie}, {Melhuish}, {Memari}, {Murphy},
  {Orlando}, {Schwarz}, {Sullivan}, {Piccirillo}, {Pryke}, {Rajguru},
  {Rusholme}, {Taylor}, {Thompson}, {Tucker}, {Turner}, {Wu}, \&
  {Zemcov}}]{Hinderks2009}
{Hinderks}, J.~R., {et~al.} 2009, \apj, 692, 1221

\bibitem[{{Hivon} {et~al.}(2002){Hivon}, {G{\'o}rski}, {Netterfield}, {Crill},
  {Prunet}, \& {Hansen}}]{Hivon2002}
{Hivon}, E., {G{\'o}rski}, K.~M., {Netterfield}, C.~B., {Crill}, B.~P.,
  {Prunet}, S., \& {Hansen}, F. 2002, \apj, 567, 2

\bibitem[{{Israel}(1998)}]{Israel1998}
{Israel}, F.~P. 1998, \aapr, 8, 237

\bibitem[{{Israel} {et~al.}(2008){Israel}, {Raban}, {Booth}, \&
  {Rantakyr{\"o}}}]{Israel2008}
{Israel}, F.~P., {Raban}, D., {Booth}, R.~S., \& {Rantakyr{\"o}}, F.~T. 2008,
  \aap, 483, 741

\bibitem[{{Leitch} {et~al.}(2002){Leitch}, {Kovac}, {Pryke}, {Carlstrom},
  {Halverson}, {Holzapfel}, {Dragovan}, {Reddall}, \& {Sandberg}}]{Leitch2002}
{Leitch}, E.~M., {et~al.} 2002, \nat, 420, 763

\bibitem[{{Lucy}(1974)}]{Lucy1974}
{Lucy}, L.~B. 1974, \aj, 79, 745

\bibitem[{{Pryke} {et~al.}(2009){Pryke}, {Ade}, {Bock}, {Bowden}, {Brown},
  {Cahill}, {Castro}, {Church}, {Culverhouse}, {Friedman}, {Ganga}, {Gear},
  {Gupta}, {Hinderks}, {Kovac}, {Lange}, {Leitch}, {Melhuish}, {Memari},
  {Murphy}, {Orlando}, {Schwarz}, {Sullivan}, {Piccirillo}, {Rajguru},
  {Rusholme}, {Taylor}, {Thompson}, {Turner}, {Wu}, \& {Zemcov}}]{Pryke2009}
{Pryke}, C., {et~al.} 2009, \apj, 692, 1247

\bibitem[{{Richardson}(1972)}]{Richardson1972}
{Richardson}, W.~H. 1972, Journal of the Optical Society of America
  (1917-1983), 62, 55

\bibitem[{{Wright} {et~al.}(2009){Wright}, {Chen}, {Odegard}, {Bennett},
  {Hill}, {Hinshaw}, {Jarosik}, {Komatsu}, {Nolta}, {Page}, {Spergel},
  {Weiland}, {Wollack}, {Dunkley}, {Gold}, {Halpern}, {Kogut}, {Larson},
  {Limon}, {Meyer}, \& {Tucker}}]{Wright2009}
{Wright}, E.~L., {et~al.} 2009, \apjs, 180, 283

\end{thebibliography}

\end{document}